\documentclass[twocolumn,aps,prl,superscriptaddress,showpacs,secnumroman,showkeys]{revtex4}
\usepackage{amsmath,bm}%
\usepackage{graphicx}%
\usepackage{epsf,epsfig,latexsym}
\begin{document}
\title{Laboratory Tests of Low Density Astrophysical Equations of State}  
\author{L. Qin}
\affiliation{Cyclotron Institute, Texas A$\&$M University, College Station, Texas 77843}
\author{K. Hagel}
\affiliation{Cyclotron Institute, Texas A$\&$M University, College Station, Texas 77843}
\author{R. Wada}
\affiliation{Institute of Modern Physics HIRFL, Chinese Academy of Sciences, Lanzhou, 730000, China}
\affiliation{Cyclotron Institute, Texas A$\&$M University, College Station, Texas 77843}
\author{J. B. Natowitz}
\affiliation{Cyclotron Institute, Texas A$\&$M University, College Station, Texas 77843}
\author{S. Shlomo}
\affiliation{Cyclotron Institute, Texas A$\&$M University, College Station, Texas 77843}
\author{A. Bonasera}
\affiliation{Cyclotron Institute, Texas A$\&$M University, College Station, Texas 77843}
\affiliation{Laboratori Nazionali del Sud, INFN, via Santa Sofia, 62, 95123 Catania, Italy}
\author{G. R\"opke}
\affiliation{University of Rostock, FB Physik, Rostock, Germany }
\author{S. Typel}
\affiliation{GSI Helmholtzzentrum f¨ur Schwerionenforschung GmbH,Theorie, 
Planckstra\ss e 1, D-64291 Darmstadt, Germany}
\author{Z. Chen}
\affiliation{Institute of Modern Physics HIRFL, Chinese Academy of Sciences, Lanzhou, 730000,China.}
\author{M. Huang}
\affiliation{Institute of Modern Physics HIRFL, Chinese Academy of Sciences, Lanzhou, 730000,China.}
\author{J. Wang}
\affiliation{Institute of Modern Physics HIRFL, Chinese Academy of Sciences, Lanzhou, 730000,China.}
\author{H. Zheng}
\affiliation{Cyclotron Institute, Texas A$\&$M University, College Station, Texas 77843}
\author{S. Kowalski}
\affiliation{Institute of Physics, Silesia University, Katowice, Poland.}
\author{M. Barbui}
\affiliation{Cyclotron Institute, Texas A$\&$M University, College Station, Texas 77843}
\author{M. R. D. Rodrigues}
\affiliation{Cyclotron Institute, Texas A$\&$M University, College Station, Texas 77843}
\author{K. Schmidt}
\affiliation{Cyclotron Institute, Texas A$\&$M University, College Station, Texas 77843}
\author{D. Fabris}
\affiliation{Dipartamento di Fisica dell~Universita di Padova and INFN Sezione di Padova, Padova, Italy}
\author{M. Lunardon}
\affiliation{Dipartamento di Fisica dell~Universita di Padova and INFN Sezione di Padova, Padova, Italy}
\author{S. Moretto}
\affiliation{Dipartamento di Fisica dell~Universita di Padova and INFN Sezione di Padova, Padova, Italy}
\author{G. Nebbia}
\affiliation{Dipartamento di Fisica dell~Universita di Padova and INFN Sezione di Padova, Padova, Italy}
\author{S. Pesente}
\affiliation{Dipartamento di Fisica dell~Universita di Padova and INFN Sezione di Padova, Padova, Italy}
\author{V. Rizzi}
\affiliation{Dipartamento di Fisica dell~Universita di Padova and INFN Sezione di Padova, Padova, Italy}
\author{G. Viesti}
\affiliation{Dipartamento di Fisica dell~Universita di Padova and INFN Sezione di Padova, Padova, Italy}
\author{M. Cinausero}
\affiliation{INFN Laboratori Nazionali di Legnaro, Legnaro, Italy}
\author{G. Prete}
\affiliation{INFN Laboratori Nazionali di Legnaro, Legnaro, Italy}
\author{T. Keutgen}
\affiliation{FNRS and IPN, Universit\'e Catholique de Louvain, B-1348 Louvain-Neuve, Belgium}
\author{Y. El Masri}
\affiliation{FNRS and IPN, Universit\'e Catholique de Louvain, B-1348 Louvain-Neuve, Belgium}
\author{Z. Majka}
\affiliation{Smoluchowski Institute of Physics, Jagiellonian University, Krakow, Poland}

\date{\today}

\begin{abstract}
Clustering in low density nuclear matter has been investigated using 
the NIMROD multi-detector at Texas A\&M University.  Thermal coalescence 
modes were employed to extract densities, $\rho$, and temperatures, $T$, 
for evolving systems formed in collisions of 47 $A$ MeV  $^{40}$Ar + 
$^{112}$Sn ,$^{124}$Sn  and  $^{64}$Zn + $^{112}$Sn , $^{124}$Sn. The 
yields of $d$, $t$, $^{3}$He, and $^{4}$He have been determined at   
$\rho$ = 0.002 to 0.03 nucleons/fm$^{3}$ and $T$= 5 to 11 MeV.  The 
experimentally derived equilibrium constants for $\alpha$ particle production 
are compared with those predicted by a number of astrophysical equations of 
state.  The data provide important new constraints on the model calculations.
\end{abstract}

\pacs{25.70.Pq}

\keywords{Intermediate heavy ion reactions, chemical equilibrium, neutron and proton chemical potential, quantum statisitcal model calculations}

\maketitle
 
\section*{I. INTRODUCTION}
Reliable understanding of the nuclear equation of state, EOS, over a 
wide range of densities and temperatures is crucial to both nuclear 
science and to our understanding of stellar evolution and 
supernovae~\cite{li08}.  In the latter context it is well known that a valid 
treatment of the correlations and clusterization in low density matter is a 
vital ingredient of astrophysical models. To meet the need  for the nuclear 
input, some extensive well known calculations and existing tabulations, 
based on varying effective interactions, were developed and have 
served as standard input for a wide variety of astrophysical 
simulations~\cite{shen98,lattimer01,sumiyoshi05}.  More recently some 
new approaches have led to a number of new 
predictions~\cite{horowitz06,sumiyoshi08,typel10,hempel10,shen11_1,shen11_2}. 
While all of the models in use lead to strong alpha clustering of the matter 
at low nucleon densities, $\rho$, and temperatures, $T$, they differ 
significantly in their quantitative predictions, usually tabulated as alpha 
mass fractions, at specified $T$, $\rho$ and proton mass fraction, $Y_{P}$. 
See references~\cite{typel10,shen11_1} for example. For a given $Y_{P}$ the 
differences in the absolute values reflect differences in the effective 
interactions inherent in the chosen nuclear equation of state, 
differences in the formulation and approximations employed in the 
models, and differences in the treatment of possible competing 
species. We present here results of an experimental determination 
of clustering yields in low density nuclear matter and use these 
results to make a direct test of the different models, focusing on 
alpha clustering.  Our test observable is not the alpha mass fraction 
but rather the equilibrium constant for alpha particle clustering. 
The model derived equilibrium constants should be nearly independent of 
proton fraction and the choice of competing species assumed in a 
particular model.

\section{Experimental Techniques}
We reported in Refs.~\cite{kowalski07} and~\cite{natowitz10} that measurements 
of nucleon and light cluster emission from the participant matter which is 
produced in near Fermi energy heavy ion collisions could be employed to probe 
the EOS at low densities and moderate temperatures where clustering is 
important.  Our data demonstrated a large degree of alpha clustering for 
densities at and below $\sim0.05$ times normal nuclear density and 
temperatures of 4 to 10 MeV.  Using these data we derived experimental 
symmetry free energies in low density nuclear 
matter~\cite{kowalski07,natowitz10}.  That analysis employed 
the isoscaling technique which compares yields for two systems with 
similar temperatures but different $N/Z$ ratios to determine the 
differences in chemical potentials and symmetry energy~\cite{tsang01,souza08}.

The NIMROD 4$\pi$ multi-detector at Texas A\&M University has now 
been used to extend our clustered matter measurements to higher densities. 
Cluster production in collisions of 47$A$ MeV  
$^{ 40}$Ar with $^{112}$$^{,}$$^{124}$Sn and  $^{64}$Zn with 
$^{112}$$^{,}$$^{124}$Sn was studied. NIMROD consists of a 166 
segment charged particle array set inside a neutron ball~\cite{wuenschel09}. 
The charged particle array is arranged in 12 rings of Si-CsI telescopes 
or single CsI detectors concentric around the beam axis. The CsI 
detectors are 1-10 cm thick Tl doped crystals read by photomultiplier 
tubes. A pulse shape discrimination method is employed to identify 
light particles in the CsI detectors. For this experiment each of 
the forward rings included two segments having two Si detectors 
(150 and 500 $\mu$m thick) in front of the CsI detectors 
(super telescopes) and three having one Si detector (300 $\mu$m thick). 
Each super telescope was further divided into two sections. Neutron 
multiplicity was measured with the 4$\pi$ neutron detector surrounding 
the charged particle array. This detector is a neutron calorimeter 
filled with Gd doped pseudocumene. Thermalization and 
capture of emitted neutrons in the ball leads to scintillation which 
is observed with phototubes providing event by event determinations 
of neutron multiplicity.  Further details on the detection system, 
energy calibrations and neutron ball efficiency may be found in 
Ref.~\cite{qin08}.  The combined neutron and charged particle 
multiplicities were employed to select the most violent events for 
subsequent analysis. 

\section{Analysis}
The dynamics of the collision process allow us to probe the nature of the 
intermediate velocity ``nucleon-nucleon'' emission 
source~\cite{hagel00,wada89,wada04}. Measurement of emission cross sections 
of nucleons and light clusters together with suitable application of a 
coalescence ansatz~\cite{mekjian78} provides the means to probe the 
properties and evolution of the interaction region.  The techniques used 
have been detailed in several previous 
publications~\cite{kowalski07,natowitz10,hagel00,wada89,wada04} and are 
described briefly below.  A notable difference from 
Refs.~\cite{kowalski07,natowitz10} is the method of density extraction.  This 
is discussed more extensively in the following.  We emphasize that the event 
election is on the more violent collisions. Cross section weighting favors 
mid-range impact parameters.

An initial estimation of emission multiplicities at each stage of the 
reaction was made by fitting the observed light particle spectra assuming 
contributions from three sources, a projectile-like fragment (PLF) 
source, an intermediate velocity (IV) source, and a 
target-like fragment (TLF) source. A reasonable reproduction of 
the observed spectra is achieved. Except for the most forward detector 
rings the data are dominated by particles associated with the 
IV and TLF sources. The IV source 
velocities are very close to 50\% of the beam velocity as seen in many 
other studies~\cite{hagel00,wada89,wada04}.  The observed spectral slopes 
reflect the evolution dynamics of the source~\cite{hagel00,bauer95,zheng11}.  
For further analysis, this IV source is most easily sampled at the 
intermediate angles where contributions from the other sources are minimized. 
For the analysis of the evolution of the source we have selected the 
data in ring 9 of the NIMROD detector. This ring covered an angular 
range in the laboratory of 38$^\circ$ to  52$^\circ$. Inspection of  invariant 
velocity plots constructed for each ejectile and each system, as 
well as of the results of the three-source fit analyses indicate 
that this selection of angular range minimizes contributions 
from secondary evaporative decay of projectile like or target like 
sources. We treat the IV source as a nascent fireball created in the
 participant interaction zone. 

The expansion and cooling of this zone leads to a correlated evolution 
of density and temperature which we probe using particle and cluster 
observables, yield, energy and angle.  As in the  previous 
work~\cite{kowalski07,natowitz10} we have employed double isotope yield 
ratios~\cite{albergo85,kolomiets97} to characterize the temperature at a 
particular emission time.  Model studies comparing Albergo 
model~\cite{albergo85} temperatures and densities to the 
known input values have shown the double isotope ratio temperatures 
to be relatively robust in this density range~\cite{shlomo09}.  
However the densities 
extracted using the Albergo model are useful only at the very lowest 
densities~\cite{shlomo09}.  Consequently, in this study we have employed 
a different means of density extraction, the thermal coalescence model of 
Mekjian~\cite{hagel00,mekjian78}.

To determine the coalescence parameter $P_0$, the radius in momentum space, 
from our data we have followed the Coulomb corrected coalescence model 
formalism of Awes \textit{et al.}`\cite{awes81} and previously employed by us 
in Ref.~\cite{hagel00}. In the laboratory frame the derived relationship 
between the observed cluster and proton differential cross sections is   

\begin{eqnarray}
\frac{d^2N(Z,N,E_A)}{dE_Ad\Omega} &=& 
R_{np}^N \frac{A^{-1}}{N!Z!} \left(\frac{4\pi P_0^3}{3[2m^3(E-E_C)]
^{1/2}}\right)^{A-1}\nonumber\\
&&\times\left(\frac{d^2N(1,0,E)}{dE\,d\Omega}\right)^A
\label{eqnCoal}
\end{eqnarray}
where the double differential multiplicity for a cluster of mass number 
$A$ containing $Z$ protons and $N$ neutrons and having a Coulomb-corrected 
energy $E_A$, is related to the proton double differential multiplicity 
at the same Coulomb corrected energy per nucleon, $E-E_C$, where $E_{C}$
is the Coulomb barrier for proton emission.  
$R_{np}$ is the neutron to proton ratio.

\begin{figure}
\epsfig{file=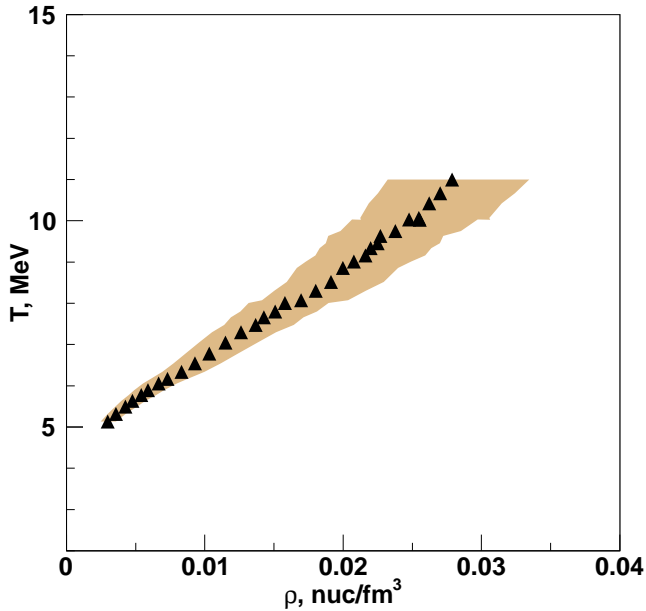,width=9.2cm,angle=0}
 \caption{Temperatures and densities sampled by the expanding IV source. The
shaded band indicates the uncertainity in density.  See text.}
\label{fig1}
\end{figure}

A strict quantitative application of the coalescence model requires 
knowledge of cluster, neutron and proton differential cross sections 
with proper absolute normalizations. In this work absolute measured 
multiplicities for the selected violent events are employed. The 
neutron spectra are not measured. However, since within the framework 
of the coalescence model the yield ratios of two isotopes which 
differ by one neutron are determined by their binding  energies and 
the $n/p$ ratio in the coalescence volume,
we have used the observed triton to $^{3}$He yield ratio to derive 
the $n/p$ ratio used in this analysis.  

In the Mekjian model thermal and chemical equilibrium determines 
coalescence yields of all species. Under these assumptions there is 
a direct relationship between the derived radius in momentum space 
and the volume of the emitting system. In terms of the $P_0$ derived 
from Eq. (\ref{eqnCoal}) and assuming a spherical source,
\begin{equation}
V=\frac{3h^3}{4\pi P_0^3}\,\,
\left[(2s+1)\left(\frac{Z!N!A^3}{2^A}\right)e^{\frac{E_0}{T}}\right]^{\frac{1}{(A-1)}}
\label{eqn2}
\end{equation}
where $h$ is Plancks constant and $Z$, $N$, and $A$ are the same as in 
Eq. (\ref{eqnCoal}).  $E_0$ is the binding energy, $s$ the spin of the emitted 
cluster, and $T$ is the temperature. Thus the radius, and therefore the 
volume, can be derived from the observed $P_0$ and temperature values assuming 
a spherical shape.  We  note that this volume is a free volume.

\begin{figure}
\epsfig{file=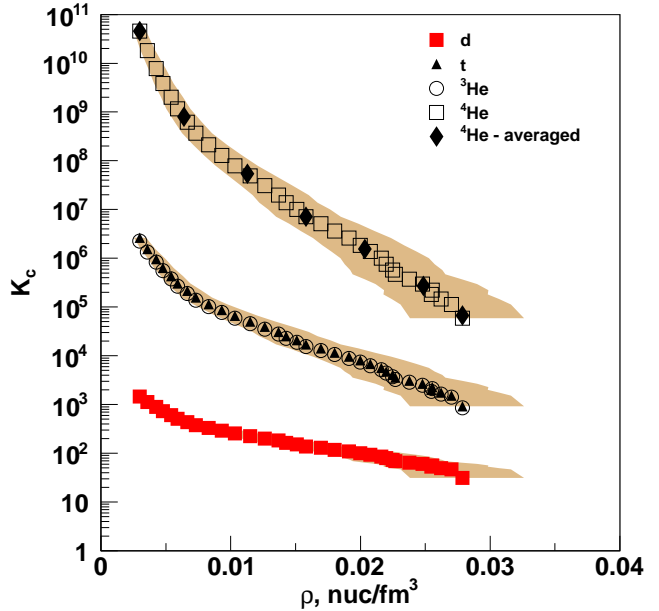,width=9.2cm,angle=0}
\caption{Equilibrium constants $K_c$ derived from the experimental yields. 
See text. Solid squares- $K_c$ (d), Solid triangles- $K_c$ (t), 
Open circles- $K_c$ ($^3$He) and Open squares $K_c$ ($\alpha$). 
Solid diamonds indicate the values of $K_c$ ($\alpha$) corresponding 
to intergral values of the temperature from 5 to 11 MeV (left to right).  The
shaded bands indicate the uncertainity in density.  See text.
}
\label{fig2}
\end{figure}

Because our goal was to derive information on the density and 
temperature evolution of the emitting system, our analysis was not 
limited to determining an average $P_0$ value.   Instead, 
as in our previous studies~\cite{hagel00}, results for $d$, $t$, 
$^3$He, and $^4$He, were derived as a function of 
$v_{\rm surf}$, the velocity of the emerging particle
 at the nuclear surface, prior to Coulomb acceleration. From the 
relevant $P_0$ values we then determined volumes using equation (\ref{eqn2}). 
A comparison of these volumes indicated good agreement for $t$, 
$^{3}$He and $^{4}$He. The volumes  derived from the deuteron data 
are typically somewhat smaller. This appears to reflect the fragility 
of the deuteron and its survival probability once formed~\cite{cervesato92}. 
For this reason we have used average volumes derived from the $A>2$
clusters in the analysis.  Given that mass is removed from 
the system during the evolution, we determined the relevant masses for 
each volume by assuming that the initial mass of the source was that 
obtained in the source fitting analysis and then deducing the 
mass remaining at a given $v_{\rm surf}$ from the observed energy spectra. 
This is also an averaging process and ignores fluctuations.  Once these
masses were known they were used to determine an excluded volume for the
particles.  Addition of this excluded volume to the free volume produced 
the total volumes needed for the density calculations.  These were determined 
by dividing the mass remaining by the total volume.  This was done as a 
function of $v_{\rm surf}$. 

\section{Results}
\subsection{Temperatures and Densities}
Inspection of the results for the four different systems studied 
revealed that the temperatures, densities and equilibrium constants 
for all systems are the same within statistical uncertainties. Therefore 
we have combined them to determine the values reported in this paper. 

We present in Fig.~\ref{fig1} the experimentally derived density and 
temperature evolution of the IV source.  Densities are expressed as total 
number of nucleons (including those in clusters) per fm$^3$.  There is a 
strong correlation of increasing temperature with increasing density.

Estimated errors on these temperatures are 10\% below $\rho=0.01$ increasing 
to 15\% at $\rho\sim0.03$.  The error in the derivation of the density arises 
from the uncertainty on the volume which is dominated by the uncertainty in 
temperature and the uncertainty in source mass derived from source fitting 
to complex spectra. The estimated errors are $\pm 17$\%.  We have included in 
each figure shaded bands representing this uncertainty in density.  On the 
log plot the uncertainties in K are much smaller.  

\begin{figure} 
\epsfig{file=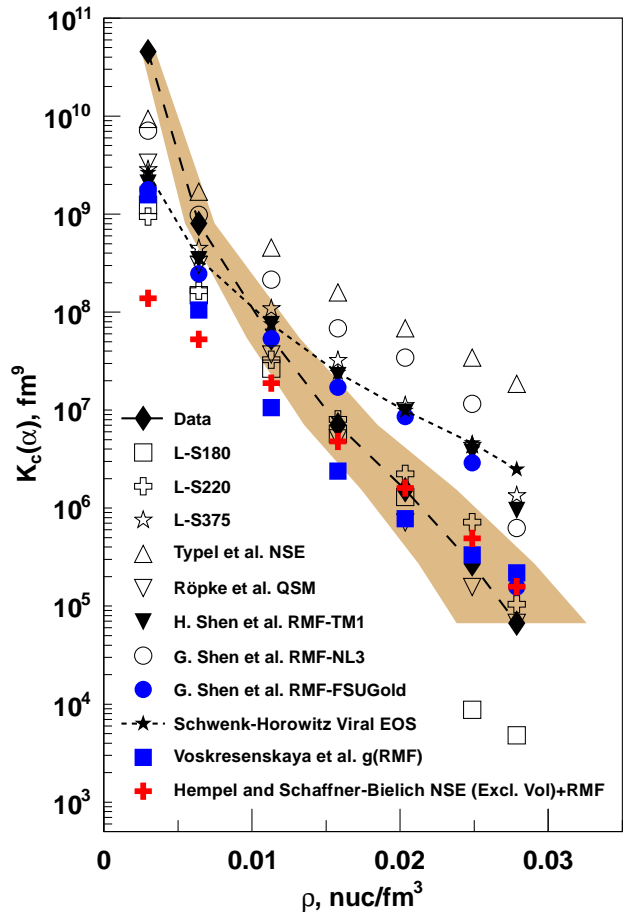,width=9.2cm,angle=0}
\caption{Comparison of experimental values of $K_c$($\alpha$) with those 
from various EOS  calculations.  See text for the theoretical models.  The 
shaded band indicates the uncertainty in density for the experimental data. 
See text.
}
\label{fig3}
\end{figure}

\subsection{Equilibrium Constants}
As already mentioned above, the absolute cluster yields and mass 
fractions calculated by the models depend upon the $N/Z$ ratio of the 
matter, the model specific nucleon-nucleon interaction assumed, and 
the various approximations of a given model. In addition, as all of 
the treatments assume chemical equilibrium, they also depend upon the 
number and type of competitive species included in the calculation.  
Historically, most EOS tables, used in astrophysical simulations, have 
treated only neutrons, protons, $\alpha$-particles, and a single 
average heavy nucleus~\cite{shen98,lattimer01}.  More recently some models 
have been developed in which the number of competing species treated has been 
greatly expanded~\cite{hempel10,shen11_1,shen11_2}.  In an equilibrium 
situation, all relevant equilibria must be simultaneously satisfied. Thus, 
if relevant species are not included, the calculated mass fractions of 
all included species will be in error. 

For this reason we do not believe that a direct comparison with 
calculated mass fractions is the appropriate way to test the individual 
models. We choose rather to compare the experimentally derived equilibrium 
constants for the production of $\alpha$ particles with those of the models. 
The model derived equilibrium constants should be independent of proton 
fraction and the choice of competing species in a particular model. 
Specifically we define the equilibrium constant, $K_{c}$, as
\begin{equation}
K_c(A,Z) = \frac{\rho(A,Z)}{\rho_p^Z\rho_n^{(A-Z)}}
\end{equation}
where $\rho(A,Z)$, $\rho_{p}$ and $\rho_{n}$ are respectively the 
densities of cluster of mass $A$ and atomic number $Z$, free protons and 
free neutrons. We express the densities in units of nucleons/fm$^{3}$.  
In Fig.~\ref{fig2} we present the equilibrium constants, $K_{c}$, 
for $d$, $t$, $^{3}$He, and $^{4}$He cluster formation as a function of 
density.  For the purpose of later comparisons with EOS calculations we have 
interpolated the experimental data for the alpha cluster to determine 
the equilibrium constants at integral temperatures from 5 to 11 MeV. 
These values are indicated by solid black diamonds on the figure. 
Temperatures increase as the density increases.  Statistical errors 
are smaller than the symbols.

For the purposes of a general assessment of the available EOS calculations 
at low density, we present in Fig.~\ref{fig3} a comparison between the 
experimentally derived values of $K_{c}$($\alpha$) and those predicted by 
available astrophysical EOS calculations. Note that this plot is  logarithmic 
in $K_{c}$. The solid black diamonds connected by a dashed line indicate the 
experimental results.  For a further comparison we show with a dotted line 
the results of the Virial EOS of Horowitz and Schwenk~\cite{horowitz06}. 
This EOS is based on virial coefficients derived from scattering data and 
has been suggested as a benchmark for other calculations at low density.  
Not surprisingly the calculated values of the equilibrium constant tend to 
converge at the lower densities.   Even at the lowest densities sampled, 
however, there are significant differences. At higher densities, 
0.01 to 0.03 nuc/fm$^{3}$, the various interactions employed all lead to a 
decrease of $K_{c}$ below that of the NSE values of 
Typel \textit{et al.}~\cite{typel10} as expected.  However there are 
significant differences in the models, both in absolute value and in the 
rate of change of $K_{c}(\alpha)$ at the higher densities.  In the density 
range above $\sim0.01$ nuc/fm$^3$ the predictions of the Virial model of 
Schwenk and Horowitz~\cite{horowitz06}, the Model of 
H. Shen \textit{et al.}~\cite{shen98}, the Lattimer Swesty Model with a Skyrme 
interaction having an incompressibility of 375 MeV~\cite{lattimer01} and those 
of G. Shen \textit{et al.} which employ the FSUGold~\cite{shen11_2} or NL3 
interaction~\cite{shen11_1} all exhibit shallower slopes than the data and 
values all well above the data. 

The model predictions showing better agreement with the data above 
0.01 nuc/fm$^3$ are the Lattimer and Swesty calculations employing Skyrme 
interactions with incompressibilities of 180 MeV and 
220 MeV~\cite{lattimer01}.  They are close to the data except for the two 
points for 180 MeV at highest density. 

The QSM model results are very close to the experimental results.  
The present calculations include the latest QSM estimates of momentum 
dependent in medium binding energy shifts~\cite{typel10,roepke11}.  Although 
well below the data and the other model calculations at the lowest velocities, 
the results from reference~\cite{hempel10}, an NSE model which includes 
excluded volume effects are in good agreement with the data at the higher 
velocities. In a recent paper the extent to which  the excluded volume 
concept can simulate the in-medium effects, in particular Pauli blocking has 
now been explored~\cite{hempel11}.

Calculations using the recently improved RMF model, taking into account 
continuum effects~\cite{typel10,voskresenskaya12}, produce predictions 
somewhat below the experimental data.  Incorporation of the latest QSM model 
derived binding energy shifts~\cite{roepke11} into this model have not yet 
been made.


While at the lowest densities the results of the theoretical calculations 
converge towards the NSE result, the experimentally derived equilibrium 
constants at these lowest densities are significantly higher.  This may 
indicate a limit in the experiment and inclusion of non-coalescence source 
of the alpha particles.  Alternatively additional contributions to the alpha 
yield resulting from quartetting correlations which favor boson condensation, 
not treated in the models, might be indicated~\cite{roepke98}.  Further 
investigations of this region might prove fruitful.  

\section{Summary and conclusions}
Calculated equilibrium constants for specific cluster formation should be 
independent of $n/p$ ratios and numbers or types of other species included. 
Even so, reported model calculations of the equilibrium constant for alpha 
formation in low density nuclear matter vary by about two orders of magnitude 
in the density and temperature region explored by the data presented in 
this paper.  These new experimental data for these equilibrium constants 
provide important constraints on the low density equations of state.  The 
use of the NSE which neglects in medium effects is clearly excluded as are 
several other models.  While it is possible that these models might be 
improved , the data strongly indicate that accounting for in-medium effects, 
as in the semi-empirical excluded volume approximation~\cite{hempel10} or 
the more sophisticated QSM approach~\cite{roepke09,roepke11}, is required. 
This is particularly evidenced in the density range of 
0.01 to 0.03 nuc/fm$^3$ in the present data.

We recall again that details of the low density composition can be important 
in modeling supernova evolution and neutron star 
properties~\cite{sumiyoshi08,hempel10,oconnor10}.  As emphasized in this 
paper, mistakes in $K_{c}$($\alpha$) translate into errors not only in the 
yields of alpha particles but also in the yields of all competing species, 
regardless of the completeness with which those other species are included. 
The errors on the equilibrium yields of heavier species will increase with 
increasing mass.


\section{Acknowledgements}
This work was supported by the United States Department of Energy under 
Grant \# DE-FG03- 93ER40773 and by The Robert A. Welch Foundation under 
Grant \# A0330.

\end{document}